\documentclass[%
reprint,
superscriptaddress,
amsmath,amssymb,
aps,
noeprint,
]{revtex4-2}
\usepackage{color}
\usepackage[utf8]{inputenc}
\usepackage{graphicx}
\usepackage{dcolumn}
\usepackage{bm}
\usepackage{upgreek}

\newcommand{\ezero}{\mbox{$E_0$}}
\newcommand{\mtwonue}{\mbox{$m^2_\nu$}}
\newcommand{\mnue}{\mbox{$m_\nu $}}

\newcommand{\mtwohad}{\mbox{$|M^2_{\rm nucl}|$}}
\newcommand{\me}{\mbox{$m_\mathrm{e}$}}
\newcommand{\bspec}{$\upbeta$-spectrum}

\newcommand{\bdecay}{$\upbeta$-decay}

\newcommand{\rbg}{$R_\mathrm{bg}$}

\newcommand{\rbeta}{$R_{\upbeta}$($E$)}
\newcommand{\nua}[1]{\ensuremath{\rlap{\kern-2.5pt\ensuremath{\overset{\scriptscriptstyle(-)}{\phantom{\nu}}}}{\ensuremath{{\nu}_{#1}}}}}

\newcommand{\RomanNumeralCaps}[1]{\MakeUppercase{\romannumeral #1}}
\usepackage{soul}
\begin{document}


\title{Bound on 3+1 active-sterile neutrino mixing
\\ from the first four-week science run of KATRIN }
\newcommand{\berlin}{Institut f\"{u}r Physik, Humboldt-Universit\"{a}t zu Berlin, Newtonstr. 15, 12489 Berlin, Germany}
\newcommand{\bonn}{Helmholtz-Institut f\"{u}r Strahlen- und Kernphysik, Rheinische Friedrich-Wilhelms-Universit\"{a}t Bonn, Nussallee 14-16, 53115 Bonn, Germany}
\newcommand{\cmu}{Department of Physics, Carnegie Mellon University, Pittsburgh, PA 15213, USA}
\newcommand{\cwru}{Department of Physics, Case Western Reserve University, Cleveland, OH 44106, USA}
\newcommand{\etp}{Institute of Experimental Particle Physics~(ETP), Karlsruhe Institute of Technology~(KIT), Wolfgang-Gaede-Str. 1, 76131 Karlsruhe, Germany}
\newcommand{\fulda}{University of Applied Sciences~(HFD)~Fulda, Leipziger Str.~123, 36037 Fulda, Germany}
%
%
%
\newcommand{\iap}{Institute for Astroparticle Physics~(IAP), Karlsruhe Institute of Technology~(KIT), Hermann-von-Helmholtz-Platz 1, 76344 Eggenstein-Leopoldshafen, Germany}
\newcommand{\ipe}{Institute for Data Processing and Electronics~(IPE), Karlsruhe Institute of Technology~(KIT), Hermann-von-Helmholtz-Platz 1, 76344 Eggenstein-Leopoldshafen, Germany}
\newcommand{\itep}{Institute for Technical Physics~(ITEP), Karlsruhe Institute of Technology~(KIT), Hermann-von-Helmholtz-Platz 1, 76344 Eggenstein-Leopoldshafen, Germany}
\newcommand{\tlk}{Tritium Laboratory Karlsruhe~(TLK), Karlsruhe Institute of Technology~(KIT), Hermann-von-Helmholtz-Platz 1, 76344 Eggenstein-Leopoldshafen, Germany}
\newcommand{\ppq}{Project, Process, and Quality Management~(PPQ), Karlsruhe Institute of Technology~(KIT), Hermann-von-Helmholtz-Platz 1, 76344 Eggenstein-Leopoldshafen, Germany    }
%
%
\newcommand{\inr}{Institute for Nuclear Research of Russian Academy of Sciences, 60th October Anniversary Prospect 7a, 117312 Moscow, Russia}
\newcommand{\lbnl}{Institute for Nuclear and Particle Astrophysics and Nuclear Science Division, Lawrence Berkeley National Laboratory, Berkeley, CA 94720, USA}
\newcommand{\madrid}{Departamento de Qu\'{i}mica F\'{i}sica Aplicada, Universidad Autonoma de Madrid, Campus de Cantoblanco, 28049 Madrid, Spain}
\newcommand{\mainz}{Institut f\"{u}r Physik, Johannes-Gutenberg-Universit\"{a}t Mainz, 55099 Mainz, Germany}
\newcommand{\mpp}{Max-Planck-Institut f\"{u}r Physik, F\"{o}hringer Ring 6, 80805 M\"{u}nchen, Germany}
\newcommand{\massit}{Laboratory for Nuclear Science, Massachusetts Institute of Technology, 77 Massachusetts Ave, Cambridge, MA 02139, USA}
\newcommand{\mpik}{Max-Planck-Institut f\"{u}r Kernphysik, Saupfercheckweg 1, 69117 Heidelberg, Germany}
\newcommand{\muenster}{Institut f\"{u}r Kernphysik, Westf\"alische Wilhelms-Universit\"{a}t M\"{u}nster, Wilhelm-Klemm-Str. 9, 48149 M\"{u}nster, Germany}
\newcommand{\npi}{Nuclear Physics Institute of the CAS, v. v. i., CZ-250 68 \v{R}e\v{z}, Czech Republic}
\newcommand{\unc}{Department of Physics and Astronomy, University of North Carolina, Chapel Hill, NC 27599, USA}
\newcommand{\washington}{Center for Experimental Nuclear Physics and Astrophysics, and Dept.~of Physics, University of Washington, Seattle, WA 98195, USA}
\newcommand{\wuppertal}{Department of Physics, Faculty of Mathematics and Natural Sciences, University of Wuppertal, Gau{\ss}str. 20, 42119 Wuppertal, Germany}
\newcommand{\saclay}{IRFU (DPhP \& APC), CEA, Universit\'{e} Paris-Saclay, 91191 Gif-sur-Yvette, France }
\newcommand{\tum}{Technische Universit\"{a}t M\"{u}nchen, James-Franck-Str. 1, 85748 Garching, Germany}
\newcommand{\tunl}{Triangle Universities Nuclear Laboratory, Durham, NC 27708, USA}
%
%
\newcommand{\ornl}{Also affiliated with Oak Ridge National Laboratory, Oak Ridge, TN 37831, USA}
%
%
%

\affiliation{\tlk}
\affiliation{\tum}
\affiliation{\saclay}
\affiliation{\ipe}
\affiliation{\etp}
\affiliation{\iap}
\affiliation{\inr}
\affiliation{\muenster}
\affiliation{\mpik}
\affiliation{\mpp}
\affiliation{\unc}
\affiliation{\tunl}
\affiliation{\madrid}
\affiliation{\wuppertal}
\affiliation{\washington}
\affiliation{\npi}
\affiliation{\itep}
\affiliation{\massit}
\affiliation{\cmu}
\affiliation{\lbnl}
\affiliation{\fulda}
\affiliation{\berlin}
\affiliation{\ppq}


\author{M.~Aker}\affiliation{\tlk}
\author{K.~Altenm\"{u}ller}\affiliation{\tum}\affiliation{\saclay}
\author{A.~Beglarian}\affiliation{\ipe}
\author{J.~Behrens}\affiliation{\etp}\affiliation{\iap}
\author{A.~Berlev}\affiliation{\inr}
\author{U.~Besserer}\affiliation{\tlk}
\author{B.~Bieringer}\affiliation{\muenster}
\author{K.~Blaum}\affiliation{\mpik}
\author{F.~Block}\affiliation{\etp}
\author{B.~Bornschein}\affiliation{\tlk}
% ...

\collaboration{KATRIN Collaboration}
\date{\today}

\begin{abstract} 
We report on the light sterile neutrino search from the first four-week science run of the KATRIN experiment in~2019. Beta-decay electrons from a high-purity gaseous molecular tritium source are analyzed by a high-resolution MAC-E filter down to 40~eV below the endpoint at 18.57~keV. We consider the framework with three active neutrinos and one sterile neutrino of mass $m_{4}$. The analysis is sensitive to a fourth mass state $m^2_{4} \lesssim$ 1000~eV$^2$ and to active-to-sterile neutrino mixing down to $|U_{e4}|^2~\gtrsim 2\cdot10^{-2}$.  No significant spectral distortion is observed
and exclusion bounds on the sterile mass and mixing are reported. These new limits supersede the Mainz results and improve the Troitsk bound for~$m^2_{4}~<$~30~eV$^2$. The reactor and gallium anomalies are constrained for $ 100~<~\Delta{m}^2_{41}~<~1000$~eV$^2$. 
\end{abstract}

\maketitle

{\em Introduction}.-- 
Over the past two decades, measurements of neutrino oscillations in the three flavour framework have determined all mixing angles and mass splittings~\cite{10.1093/ptep/ptaa104}. In 2011, a re-evaluation of the prediction of electron antineutrino spectra emitted by nuclear reactor cores revealed a significant discrepancy between detected and expected fluxes at baselines below 100 m, known as the reactor antineutrino anomaly (RAA)~\cite{PhysRevD.83.073006}. Moreover, both the GALLEX and SAGE solar-neutrino detectors had earlier independently reported a deficit of electron neutrino flux from $^{37}$Ar and $^{51}$Cr electron-capture decays~\cite{Hampel:1997fc,Abdurashitov:2009tn}, known as the gallium anomaly (GA). 
The significance of these anomalies is debated, mainly due to the difficulty of assessing systematic uncertainties (see~\cite{PhysRevD.101.015008,KOSTENSALO2019542} for recent discussions).
Nonetheless, these electron neutrino disappearance data could be explained by assuming the existence of a sterile neutrino, i.e.\ a non-standard neutrino that does not interact weakly, with a mass of the order of 1~eV or larger~\cite{Giunti_2019}. 

\begin{figure*}[t!]
\begin{center}
\includegraphics[width=15cm]{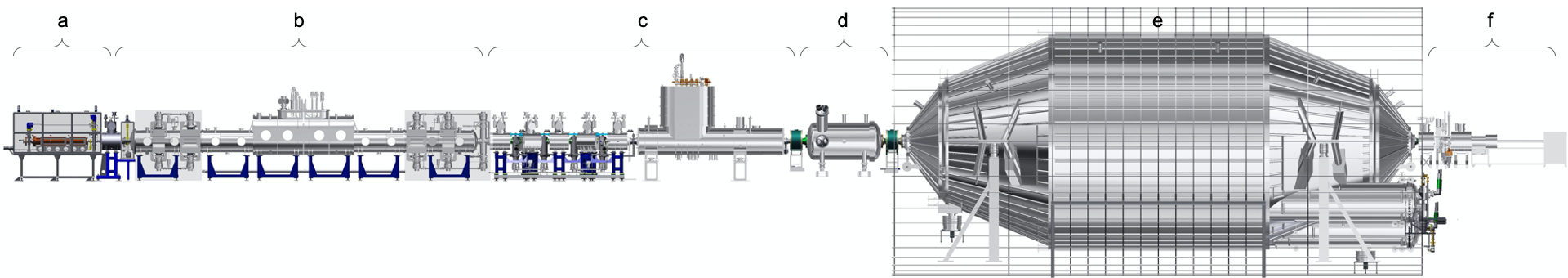}
\caption{\label{fig:katrin_setup} Main components of the KATRIN experiment: a) the rear section, b) the windowless gaseous tritium source, c) the pumping section, d) the pre-spectrometer, e) the main spectrometer, f) the electron detector.}
\end{center}
\end{figure*}

\begin{figure}[htbp]
\begin{center}
\includegraphics[width=0.43\textwidth]{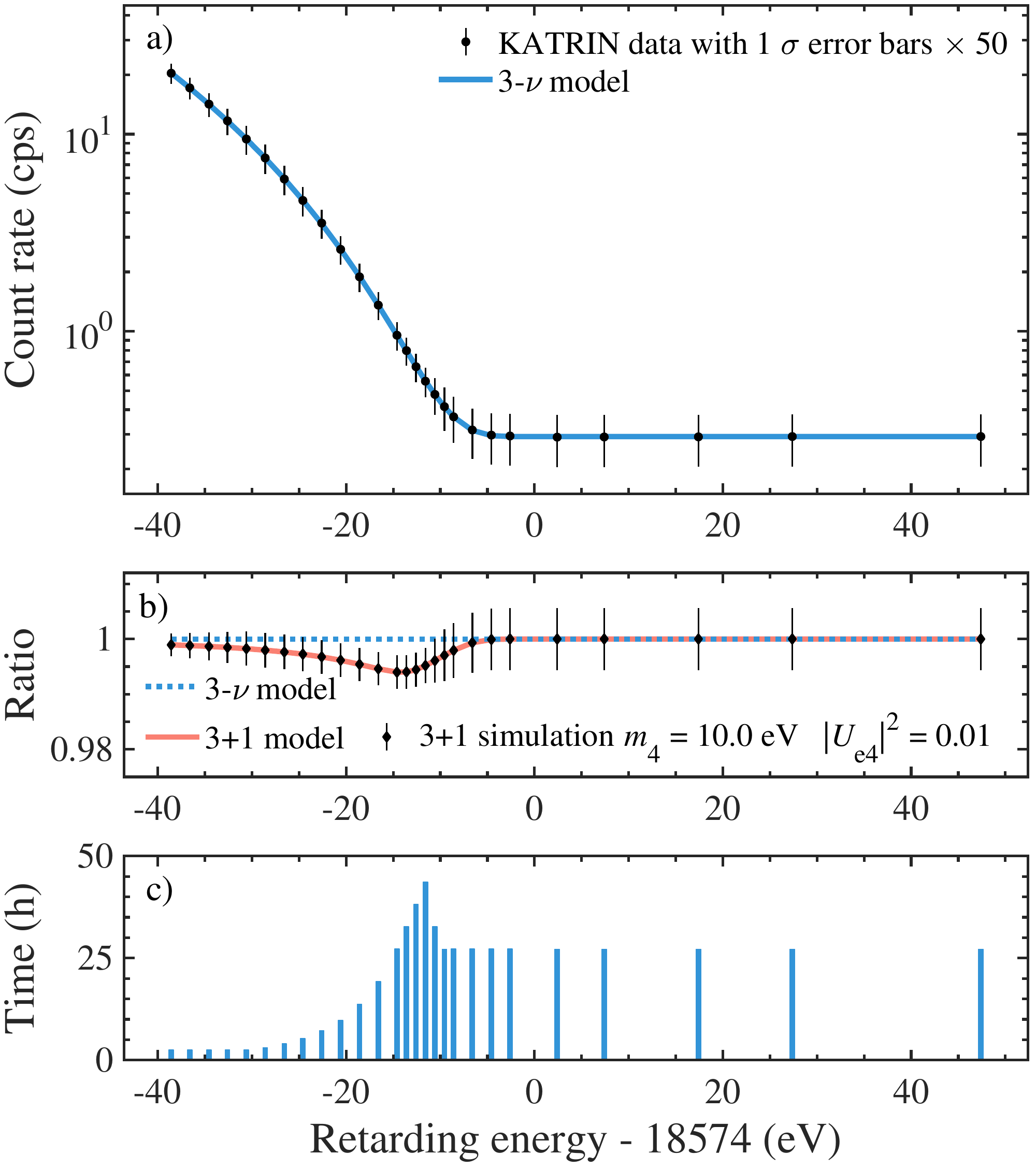}
\caption{\label{fig:beta_spectrum} a) Electron spectrum of experimental data $R(\langle qU \rangle)$ over the interval $\lbrack \ezero -$40~eV, $\ezero +$50~eV$\rbrack$ from all 274 tritium scans and the three-neutrino mixing best-fit model $R_\mathrm{calc} (\langle qU \rangle)$ (line). The integral \bdecay\ spectrum extends to \ezero\ on top of a flat background \rbg. 1-$\sigma$ statistical errors are enlarged by a factor 50. b) Simulation of an arbitrary sterile neutrino imprint on electron spectrum. The ratio of the simulated data without fluctuation, including a fourth neutrino of mass $m_4$~=~10~eV and mixing $|U_{e4}|^2$~=~0.01, to the three-neutrino mixing model is shown (red solid line). c) Integral measurement time distribution of all 27~HV set points.}
\end{center}
\end{figure}

The Karlsruhe Tritium Neutrino experiment (KATRIN)~\cite{Osipowicz:2001sq,KDR2004,Arenz:2018kma,Kleesiek:2018mel}, displayed in figure \ref{fig:katrin_setup}, provides high-precision electron spectrum measurement of the tritium $\upbeta$-decay, $^3$H $\rightarrow$ $^3$He$^+$ + e$^-$ + $\bar{\nu}_\mathrm{e}$ (endpoint \ezero\ = 18.57~keV, half-life $t_{1/2}$ = 12.32~yr). KATRIN has been designed to improve the sensitivity on the effective mass of neutrinos, \mnue, down to 0.2~eV (90$\%$ confidence level). Based on its first four-week science run in spring 2019, KATRIN reported an effective neutrino mass square value of $(-1.0~ ^{+~ 0.9}_ {-~ 1.1})$~eV$^2$, leading to an upper limit of \mnue~$<$ 1.1~eV (90$\%$ confidence level)~\cite{PhysRevLett.123.221802}. Using the same data set, it is also possible to limit the mass and mixture of a sterile neutrino. Indeed, a new  non-standard neutrino mass state, mainly sterile, would manifest itself as a distortion of the spectrum of the $\upbeta$-decay electrons. The signature would be a kink-like feature, as can be seen with a simulation presented in~figure~\ref{fig:beta_spectrum} b, standing out on top of the expected smooth tritium $\upbeta$-spectrum (figure~\ref{fig:beta_spectrum} a).
Previous phenomenological studies have examined the ultimate sensitivity of KATRIN to light sterile neutrinos~\cite{Riis:2010zm, Formaggio:2011jg, Esmaili:2012vg}. Moreover, an analysis of the four-week data set from the first science run was undertaken in \cite{Giunti:2019fcj}, although not all the inputs and data necessary for a comprehensive analysis were available yet. In this letter, we report the first result of KATRIN's search for light sterile neutrinos down to 40~eV below the endpoint.

{\em Experimental Setup}.-- KATRIN combines a windowless gaseous molecular tritium source (WGTS)~\cite{Robertson:1991vn}, with a spectrometer section based on the principle of magnetic adiabatic collimation with electrostatic filtering (MAC-E-filter)~\cite{Lobashev:1985mu, Picard1992,Kraus2005,Aseev2011}. 
Figure~\ref{fig:katrin_setup} gives an overview of the 70~m long experimental setup located at the Karlsruhe Institute of Technology (KIT) in Germany. Source-related components in contact with tritium, displayed in~figure~\ref{fig:katrin_setup} a-c, are  integrated into the Tritium Laboratory Karlsruhe to enable a closed cycle of tritium~\cite{Priester:2015bfa}. High-purity tritium gas is continuously injected at 30~K into the WGTS. The gas then diffuses to the ends of the source where it is pumped out by a series of turbomolecular pumps. In combination with the 3~K cryo\-trap (CPS), the flow rate of tritium into the following spectrometer and detector section (figure~\ref{fig:katrin_setup} d-f) is reduced by more than 14~orders of magnitude to suppress source-related background~\cite{Osipowicz:2001sq}. 
Electrons are adiabatically guided towards the spectrometers by the source magnetic field (\emph{B}$_\mathrm{WGTS}$ = 2.52~T) and other superconducting magnets~\cite{Arenz:2018jpa} in the pumping section. 
High-precision spectroscopy is achieved by the MAC-E-filter technique, where electrons of charge $q$ are guided by the magnetic field and filtered by an electrostatic barrier called the retarding potential energy, $qU$, set by a specific High Voltage (HV) setting.
Only electrons with enough energy to overcome this barrier are transmitted to the detector section. By varying and precisely monitoring $qU$ the tritium $\upbeta$-decay spectrum is scanned in an integral mode, with an energy resolution  $\Delta E = 2.8$~eV at~\ezero. Transmitted electrons are finally counted, as a function of $qU$, in a segmented monolithic silicon detector array composed of 148 pixels~\cite{Amsbaugh:2014uca}. 

{\em Measurements of the tritium $\beta$-spectrum}.-- 
The performance of the KATRIN systems and integrated beamline, specified in~\cite{KDR2004}, was established by a sequence of long-term measurements reported in~\cite{Arenz:2018kma,Arenz:2018jpa,Priester:2015bfa,Arenz:2018ymp,Altenmuller_2020}. 
In this work we use the data from KATRIN's first high-purity tritium campaign, which ran from April 10 to May 13, 2019, at an average source activity of $2.45\cdot10^{10}$ Bq. The averaged column density $\rho d_{\mathrm{exp}}$ = $1.11\cdot10^{17}$~molecules~$\mathrm{cm}^{-2}$ of our analyzed data sample is a factor of 5 less than its nominal value. The isotopic tritium purity $\varepsilon_{\mathrm{T}}$ (0.976) is derived from the average concentration of the tritiated species T$_2$ (0.953), HT (0.035), and DT (0.011), continuously monitored using Raman spectroscopy~\cite{s20174827}.

The integral tritium \bdecay\ spectrum is scanned repeatedly in a range of $\lbrack \ezero -$90~eV, $\ezero +$50~eV$\rbrack$ by applying series of non-equidistant HV settings to the spectrometer electrode system. Each scan takes a net time of 2~h.
At each HV set point, the transmitted electrons are counted over time intervals varying from 17~to~576~s. In this search for sterile neutrinos, we analyze a range of scans covering the region from 40~eV below \ezero\ (22~HV set points) to 50~eV above (5~HV set points), as shown in the measurement time distribution in~figure~\ref{fig:beta_spectrum} c. 

{\em Data Analysis}.- We first apply quality cuts to slow control parameters associated with each tritium scan (27~HV set points). This results in the selection of 274~stable scans with an overall scanning time of 521.7~h. We then select the 117 most similar pixels (79\% of the sensitive area of the detector) and combine them into a single large effective pixel. The good temporal stability of the scanning process is verified by fits to the 274~single scans \bdecay\ endpoints. The good temporal stability allows us to stack the data from all 274~scans into one single 90-eV-wide spectrum, which is displayed in~figure~\ref{fig:beta_spectrum}a in units of counts per second (cps). Thanks to the excellent reproducibility of the HV set points $qU_{l,k}$ over all scans $k$ for each $l = 1 - 27$ (RMS~=~34~mV), we associate the mean HV $\langle qU \rangle_l$ over all scans $k$ to each HV set point $l$.
The resulting stacked integral spectrum, 
\emph{R}($\langle qU \rangle$), includes $2.03\cdot10^6$~events, with $1.48\cdot10^6$  \bdecay\-electrons expected below \ezero\ and a flat background ensemble of $0.55\cdot10^6$~events in the whole scan interval ($0.41\cdot10^6$~background events expected below \ezero). This background originates from the spectrometer and has two major sources: first, the thermal ionization of Rydberg atoms sputtered off the inner spectrometer surfaces by $^{206}$Pb-recoil ions following $\upalpha$-decays of $^{210}$Po; second, the secondary electrons induced by $\upalpha$-decays of single $^{219}$Rn atoms emanating from the vacuum pumps of the spectrometer. The resulting electrons at sub-eV energies are accelerated to $qU$ by the MAC-E-filter. The radon-induced background causes a small non-Poissonian component (see~\cite{PhysRevLett.123.221802} for details). Nonetheless, in comparison to short baseline oscillation experiments~\cite{Danilov:2019aef, PhysRevLett.121.251802, AlmazanMolina:2019qul,
Danilov:2019aef}, our search for sterile neutrinos is performed with a high signal-to-background ratio, rapidly increasing from~1~at~$\langle qU \rangle$=\ezero$-$12~eV to $>$70 at~$\langle qU \rangle$=\ezero$-$40~eV. 

{\em Experimental spectrum modelling}.-- 
The modelled experimental spectrum $R_{\mathrm{calc}}(\langle qU \rangle)$ is the convolution of the differential \bspec\ \rbeta\  with the calculated response function $f(E - \langle qU \rangle)$, in addition to an energy-independent background rate $R_\mathrm{bg}$:
\begin{eqnarray}  
R_\mathrm{calc} (\langle qU \rangle) = \nonumber & & A_\mathrm{s}\,\cdot\,N_\mathrm{T}\int R_{\mathrm{\upbeta}}(E) \cdot  f(E - \langle qU \rangle)~ dE \\ 
& &  + R_\mathrm{bg}~,
\end{eqnarray}
where $A_{\mathrm{s}}$ is the tritium signal amplitude. $N_\mathrm{T}$ denotes the number of tritium atoms in the source multiplied with the accepted solid angle of the setup $\Delta \Omega/4\pi = (1- \cos{ \theta_\mathrm{max}})/2$, with $\theta_{\mathrm{max}} = 50.4^{\circ}$,  and the detector efficiency (0.95).

The differential spectrum \rbeta\ from the super-allowed \bdecay\ of molecular tritium is given by: 
\begin{eqnarray}
\label{eq2}
   R_\upbeta(E) 
   = & & \frac{G_\mathrm{F}^2 \cdot \cos^2\Theta_\mathrm{C}}{2 \pi^3} \cdot \mtwohad \cdot F(E ,Z') \\ \nonumber
             &\cdot &  (E + \me ) \cdot \sqrt{(E + \me)^2 - \me^2} \\ \nonumber
    &\cdot & \sum_{\rm j} \zeta_j \cdot \varepsilon_j \cdot \sqrt{\varepsilon_{j}^2 - \mtwonue} \cdot \Theta(\varepsilon_{j}  - \mnue) ~ ,
\end{eqnarray}
with the square of the energy-independent nuclear matrix element $\mtwohad$, the neutrino energy $\varepsilon_j = \ezero - E - V_j$, the Fermi constant $G_\mathrm{F}$, the Cabibbo angle $\Theta_\mathrm{C}$, the electron mass $m_\mathrm{e}$, and the Fermi function $F(E,Z'=2)$. 
The calculation of \rbeta\ involves the sum over a final-state distribution (FSD) which is given by the probabilities $\zeta_j$ with which the daughter ion $^3$HeT$^+$ is left in
a molecular (i.\,e.\ a rotational, vibrational, and electronic) state with excitation energy $V_j$, as described in~\cite{PhysRevLett.123.221802}. In addition, our calculations include radiative corrections~\cite{Kleesiek:2018mel} and we take into account the thermal Doppler broadening at 30~K. 

The response function $f(E - \langle qU \rangle)$ describes the transmission probability of an electron as a function of its surplus energy $E-\langle qU \rangle$. It depends on the angular spread of electrons and the amount of neutral gas they pass through in the source, where they can undergo inelastic scattering processes (see~\cite{PhysRevLett.123.221802} for details). 

In the standard three-neutrino framework the effective neutrino mass is defined through 
$\mtwonue =
{ \sum_{k=1}^{3} |U_{ek}|^2 m_{k}^2 }$, where $U$ is the $3\times3$ {\rm PMNS} unitary mixing matrix
and $m_{k}$ is the mass of the neutrino $\nu_{k}$,
with $k=1,2,3$.
According to this framework, later referred to as the null hypothesis, the experimental spectrum \emph{R}($\langle qU \rangle$) is well described by the model of the response function $f$ and by the background $R_\mathrm{bg} = (293 \pm 1)$~mcps mainly constrained by the 5~HV set points above \ezero. A fit of $A_{\mathrm{s}}$, \ezero , $R_\mathrm{bg}$, and \mtwonue, allowed us to constrain the active neutrino mass to be less than 1.1~eV at 90$\%$ confidence level~\cite{PhysRevLett.123.221802}. 

{\em Sterile Neutrino Search}.-- 
We now extend the experimental modelling, fit and statistical analysis to constrain both the sterile neutrino mass squared $m^2_{4}$ and its mixing $|U_{e4}|^2$. In all other respects we apply the same analysis strategy as for our first neutrino mass analysis~\cite{PhysRevLett.123.221802}.

In the extension to 3+1 active-sterile neutrinos, the effective neutrino mass can be redefined as $\mtwonue=\sum |U_{ek}|^2 m_{k}^2 (1-|U_{e4}|^{2})^{-1}$. The electron spectrum, $R_\upbeta$, is replaced by $R_\upbeta(E,\mnue,m_4) = (1~-~|U_{e4}|^2) R_\upbeta(E,\mtwonue) + |U_{e4}|^2 R_\upbeta(E,m^2_{4})$, where $U$ is the extended $4\times4$ unitary matrix, $R_\upbeta(E,\mtwonue)$ is the differential electron spectrum  (eq. \ref{eq2}) associated to decays including active neutrinos in the final state, and $R_\upbeta(E,m^2_{4})$ describes the additional spectrum associated to decays involving a sterile neutrino of mass $m_{4}$. 
The observable integral spectrum $R_\mathrm{calc}$ is henceforth modeled with six free parameters: the signal amplitude $A_{\mathrm{s}}$, the effective \bdecay\ endpoint \ezero , the background rate $R_\mathrm{bg}$, the effective active neutrino mass squared \mtwonue, the sterile neutrino mass squared $m_{4}^2$, and the mixing parameter $|U_{e4}|^2$. This extended model $R_{\mathrm{calc}}(\langle qU \rangle)$ is then fitted to the experimental data $R(\langle qU \rangle)$. 

In order to mitigate bias, the full analysis is first conducted on a Monte Carlo-based (MC) data set before turning to the actual data. For each experimental scan $k$ we generate a ``MC twin'', $R_\mathrm{calc}$($\langle qU \rangle$)$_k$, from its averaged slow-control parameters and the measured background rate and endpoint.
The ``MC twin" analysis allows us to verify the accuracy of our parameter inference by retrieving the correct input MC values. This approach is also used to calculate the expected sensitivity and to assess the impact of each systematic uncertainty, described in detail in \cite{PhysRevLett.123.221802}.

The fit of the experimental spectrum $R(\langle qU \rangle)$ with the extended model $R_\mathrm{calc} (\langle qU \rangle)$ is performed by minimizing the standard $\chi^2$-estimator. In view of conducting a ``shape-only'' fit, both \ezero\ and $A_{\mathrm{s}}$ are left unconstrained.
To propagate systematic uncertainties, a covariance matrix is computed after performing ${\cal O}(10^4)$ simulations of $R_\mathrm{calc} (\langle qU \rangle)$, while varying the relevant parameters for each calculation according to their likelihood~\cite{Barlow:213033,DAgostini:1993arp,Aker:2019qfn, PhysRevLett.123.221802}. The sum of all matrices encodes the total uncertainties of $R_\mathrm{calc} (\langle qU \rangle)$ and their HV set point-dependent correlations. 
The $\chi^2$-estimator is then minimized to determine the best-fit parameters, and the shape of the $\chi^2$-function is used to infer the uncertainties. 

To obtain the sterile neutrino constraints, fits are performed on a 50 $\times$ 50 $\lbrack$ log($|U_{e4}|^2$), log($m^2_{4}$) $\rbrack$ grid (starting at the null hypothesis), by keeping $|U_{e4}|^2$ and $m^2_{4}$ constant and minimizing the $\chi^2$ with respect to all other free parameters. A finer mesh does not significantly change our results. The 95\% confidence level is given by the contour, where $\Delta\chi^{2} = \chi^2 - \chi^2_\mathrm{min} = 5.99$, assuming Wilk’s theorem~\cite{Wilks:1938dza} for two degrees of freedom. $\chi^2_\mathrm{min}$ is the global minimum of all $\chi^2$ values obtained in the grid scan. We have verified that the global minimum lies within the physical region defined as $|U_{e4}|^2 \in [0, 0.5]$ and $m^2_{4} \ge 0$. 
The correct coverage of this approach is validated by means of MC simulations, where thousands of experiments were generated for the null hypothesis and a few sterile neutrino signal hypotheses and analyzed in turn. 

{\em Results}.-
The actual fit range $\lbrack \ezero -$40~eV, $\ezero +$50~eV$\rbrack$ is such that statistical uncertainties on $ |U_{e4}|^2 $ fully dominate over systematic uncertainties in the whole range of $m_{4}^2$ considered~\cite{PhysRevLett.123.221802,KNM1long}. 

In our first and main analysis, labelled case~\RomanNumeralCaps{1}, we consider the hierarchical scenario $m_{1,2,3} \ll m_{4}$, which justifies setting $\mnue$ to its minimum allowable value. In this work we set $\mnue$ to zero, which is consistent within our sensitivity with the lower limit derived from neutrino oscillations (0.009~eV, see~\cite{10.1093/ptep/ptaa104}). 

\begin{figure}[htbp]
\begin{center}
\includegraphics[width=0.42\textwidth]{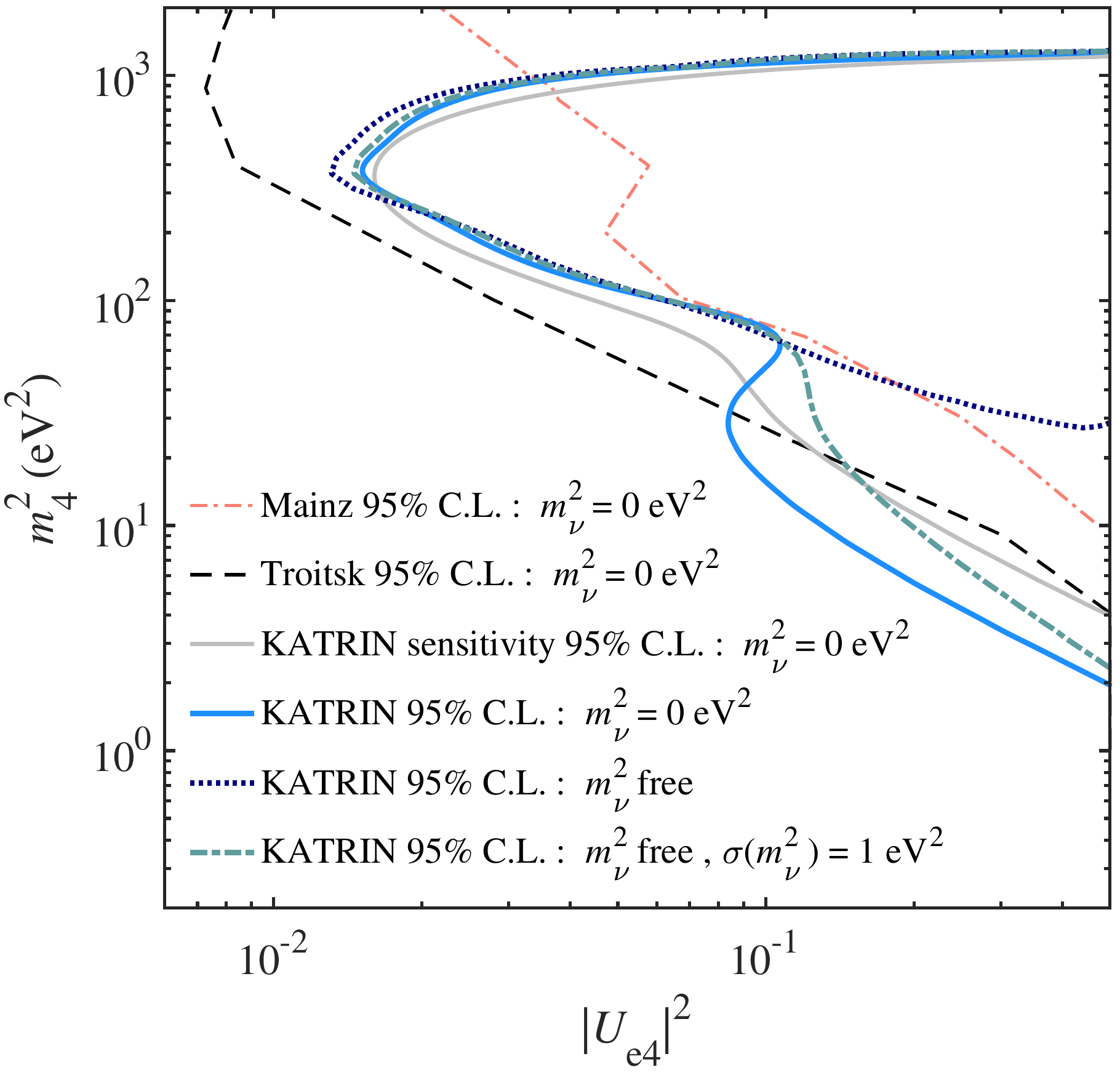}
\caption{\label{fig:contourm4sinthetasq} 
95\%~confidence level exclusion curves in the
($|U_{e4}|^2$, $m^2_{4}$) plane obtained from the analysis of KATRIN data including statistical and systematic uncertainties. The two solid lines show the expected sensitivity (light grey) and the associated exclusion (blue) for fixed $\mtwonue=0$~eV$^2$ (case~\RomanNumeralCaps{1}). The dotted line in dark blue illustrates the exclusion curve obtained with a free $\mtwonue$ (case~\RomanNumeralCaps{2}). Lastly, the dot-dashed line in turquoise displays the intermediate exclusion curve with a free $\mtwonue$ constrained with an uncertainty $\sigma (\mtwonue)$~=~1~eV$^2$ (case~\RomanNumeralCaps{3}).
First KATRIN results supersede the Mainz exclusion limit~\cite{Kraus:2012he} and improve Troitsk bounds~\cite{Belesev_2013} for~$m^2_{4}~<$~30~eV$^2$. 
}
\end{center}
\end{figure}

For each pair of ($|U_{e4}|^2$, $m^2_{4}$) a fit compares the experimental spectrum \emph{R}($\langle qU \rangle$) to the model $R_\mathrm{calc}(\langle qU \rangle)$ by only considering $A_\mathrm{sig}$, $R_\mathrm{bg}$, $E_{0}$, as free parameters. 
The global best fit minimum is found for the values $m_4^2$ = 73.0~eV$^2$, and $|U_{e4}|^2$ =~0.034. The $\chi^2$ difference between this best fit and the null hypothesis is $\Delta\chi_\mathrm{bf}^2$~=~1.6. 
Assuming the null hypothesis, the probability to obtain $\Delta\chi_\mathrm{bf}^2\geq 1.6$ is 50\%, based on the simulation and analysis of 2000 pseudo experiments. 
Therefore, our experimental result is consistent with the null hypothesis and there is no evidence for a sterile neutrino signal. Resulting 95\%~confidence level exclusion and sensitivity curves in the ($|U_{e4}|^2$, $m^2_{4}$) plane are shown in~figure~\ref{fig:contourm4sinthetasq}.  Our results agree well with the sensitivity estimates. Since the data cover only the last 40~eV of the $\beta$-spectrum this analysis is only sensitive to $m^2_{4} < 1600$~eV$^2$, with a maximum sensitivity for $m^2_{4} \simeq 400$~eV$^2$. For smaller $m^2_{4}$ the sensitivity decreases due to the reduction of statistics and vanishes for $m^2_{4} \simeq 2$~eV$^2$. For larger $m^2_{4}$ the sensitivity rapidly drops due to the narrow interval in which a sterile neutrino could influence the experimental $\beta$-spectrum.
Case~\RomanNumeralCaps{1} allows a direct comparison with previous and similar experiments. This work supersedes the Mainz exclusion limit~\cite{Kraus:2012he} for $m^2_4 \lesssim 1000$ ~eV$^2$ and improves the Troitsk bounds~\cite{Belesev_2013} for $m^2_4~ \lesssim~30$~eV$^2$, as displayed in~figure~\ref{fig:contourm4sinthetasq}.

In a second analysis, called case~\RomanNumeralCaps{2}, we consider a more generic scenario where $\mtwonue$ is treated as a free and and unconstrained parameter in the fit. Figure~\ref{fig:contourm4sinthetasq} shows the resulting 95\%~confidence level exclusion curves in the ($|U_{e4}|^2$, $m^2_{4}$) plane, only deviating from the upper limit of case~\RomanNumeralCaps{1} for $m^2_{4}<60$~eV$^2$. For low mixing, $|U_{e4}|^2 \lesssim 0.3$, and small $m^2_{4}<10$~eV$^2$, the $\mtwonue$ fitted values are in excellent agreement with our standard neutrino mass analysis~\cite{PhysRevLett.123.221802}.

As complementary information, called case~\RomanNumeralCaps{3}, we display in~figure~\ref{fig:contourm4sinthetasq} the intermediate exclusion curve that is obtained when $\mtwonue$ is treated as a nuisance parameter in the fit, using a Gaussian pull term with the expectation $\mtwonue$~=~0~eV$^2$ and a Gaussian uncertainty $\sigma (\mtwonue)$~=~1~eV$^2$.

\begin{figure}[htbp]
\begin{center}
\includegraphics[width=0.47\textwidth]{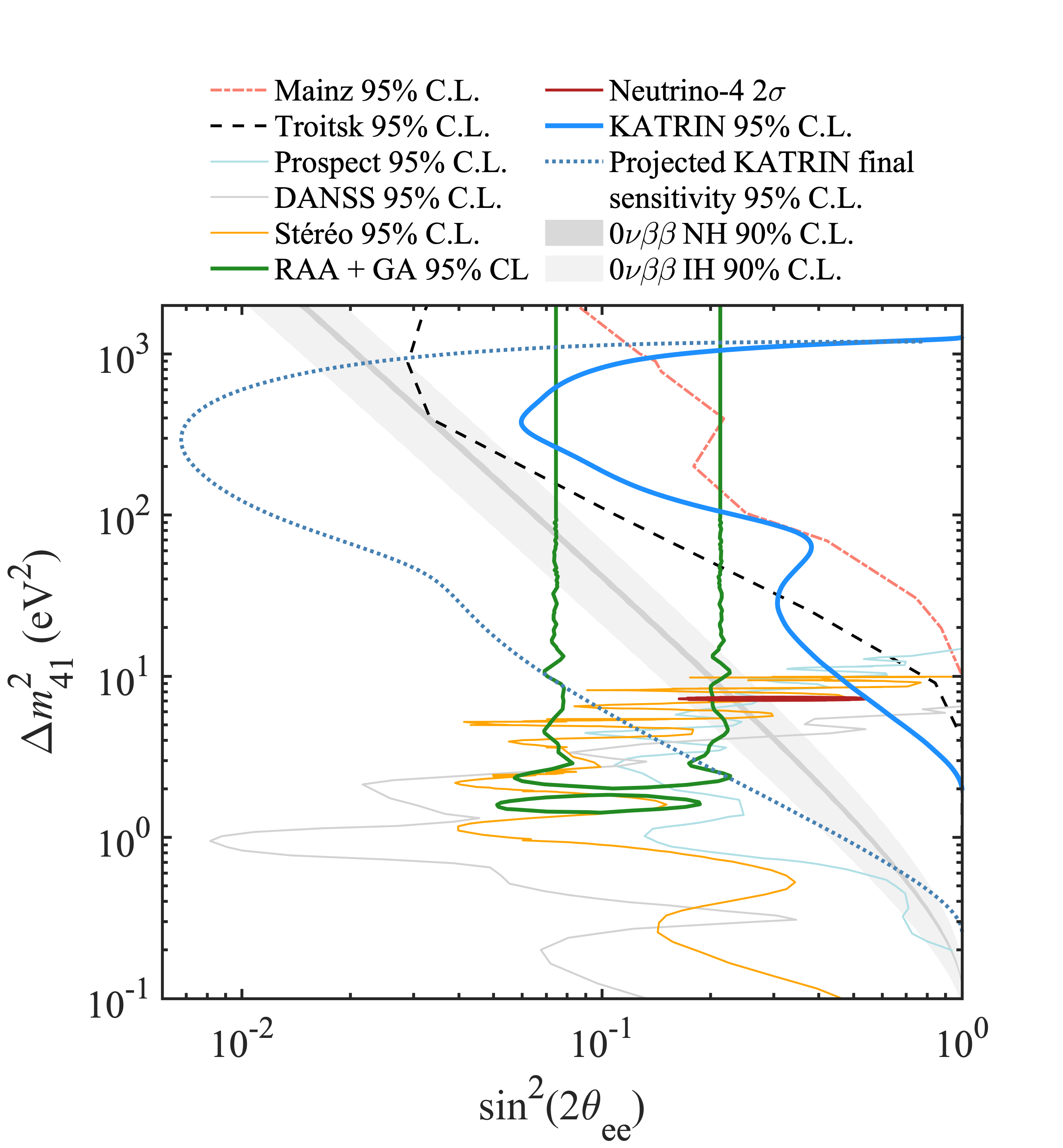}
\caption{\label{fig:contourfinal} 
95\%~confidence level exclusion curves in the
($\sin^2(2\theta_{ee}),\Delta{m}^2_{41}$)
plane obtained from the analysis of
KATRIN data with fixed $\mnue=0$.
The green contour delimits the 3+1 neutrino oscillations allowed at 95\%~confidence level by the reactor and gallium anomalies~\cite{PhysRevD.83.073006}.
KATRIN data improve the exclusion of the high $\Delta{m}^2_{41}$ values with respect to DANSS, PROSPECT, and St\'er\'eo reactor spectral ratio measurements~\cite{Danilov:2019aef, PhysRevLett.121.251802, AlmazanMolina:2019qul}. 
Mainz~\cite{Kraus:2012he} and
Troitsk~\cite{Belesev_2013} exclusion curves~\cite{Giunti_2013} are also displayed for comparison.
An estimation of KATRIN's final sensitivity (1000 days at nominal column density) is represented by the dotted line. The light (dark) gray bands delimit the exclusions from $0\nu\beta\beta$ experiments, for the case of inverted and normal hierarchies (the extension of the bands reflects the uncertainties of the parameters of the PMNS matrix~\cite{10.1093/ptep/ptaa104}).}
\end{center}
\end{figure} 
{\em Comparison with neutrino oscillation experiments}.-
At this stage it is interesting to compare our results (case~\RomanNumeralCaps{1}) with short baseline neutrino oscillation experiments measuring the electron (anti-)neutrino survival probability $P(\Delta{m}^2_{41},\sin^2(2\theta_{ee}))$~\cite{Giunti_2019}. To relate  the obtained results to KATRIN, the mass splitting can be written as $\Delta{m}^2_{41} \simeq m_4^2 - m^2_\nu$. According to the measured neutrino oscillation parameters, this approximation is valid within~$2\cdot10^{-4}$~eV$^2$, the main difference being due to the unknown mass hierarchy and the amplitude of the mixing between active and sterile neutrinos~\cite{Giunti:2019fcj}. For the analysis case~\RomanNumeralCaps{1} we simply have $\Delta{m}^2_{41} \simeq m_4^2$. Furthermore KATRIN is directly sensitive to $|U_{e4}|^2$ whereas oscillation experiments measure $\sin^2(2\theta_{ee}) = 4 |U_{e4}|^2 ( 1 - |U_{e4}|^2 )$.
These first KATRIN data exclude the high $\Delta{m}^2_{41}$ solution of the reactor and gallium anomalies, for $\Delta{m}^2_{41}$ values between 100 and 1000~eV$^2$, as depicted in~figure~\ref{fig:contourfinal}. Our data also improve the exclusion of DANSS, PROSPECT, and St\'er\'eo reactor spectral ratio measurements~\cite{Danilov:2019aef, PhysRevLett.121.251802, AlmazanMolina:2019qul} for $\Delta{m}^2_{41} \gtrsim$ 10~eV$^2$. The Neutrino-4 hint of large active-sterile mixing~\cite{Serebrov:2018vdw} is at the edge of our current 95\%~confidence level exclusion. Other sterile neutrino searches, based on $\bar{\nu}_e$ measurements at medium baseline, such as~\cite{PhysRevLett.125.071801,Yeo_2017, Abrahao:2020eta}  are sensitive to $\Delta{m}^2_{41} \lesssim$~1-10~eV$^2$ and are not displayed in~figure~\ref{fig:contourfinal}.

An estimation of KATRIN's five-year sensitivity is presented in~figure~\ref{fig:contourfinal}, assuming 1000 days of data taking at the nominal column density, the current reduced background rate of 130~mcps, and design systematic uncertainties~\cite{KDR2004}. KATRIN results will be complementary to short baseline reactor neutrino experiments, improving the global sensitivity for $\Delta{m}^2_{41} \gtrsim $~5~eV$^2$.

If sterile neutrinos with $|U_{e4}| \neq 0$ are Majorana particles, they will contribute to the effective mass $m_{\beta\beta} = |\sum U_{ei}^2 m_i|$ relevant for $0\nu\beta\beta$~\cite{PhysRevD.83.073006,Barry:2011wb,Giunti:2015kza}. Considering $m_4 \gg m_{1,2,3}$ we have the active neutrino contribution to $m_{\beta\beta}$ within the ranges 0.01 to 0.05 eV (0 to 0.005 eV) for the inverted (normal) ordering. Our current and future constraints on $U_{e4}$ and $m_4$ can then be confronted with the latest constraints of $0\nu\beta\beta$ experiments~\cite{KamLAND-Zen:2016pfg,Agostini:2020xta}, as shown in  Figure~\ref{fig:contourfinal}.

{\em Conclusion and outlook}.-- 
We have presented an initial search for signatures of a sterile neutrino admixture (3+1 framework) using the first four-week science run of the KATRIN experiment. This search comprises $1.48\cdot10^6$ \bdecay\ electrons and $0.41\cdot10^6$ background events below~\ezero, with a signal to background ratio of up to~70. The analysis is sensitive to mass values $m_4$ ranging from about 2~to~40~eV. No significant sterile neutrino signal is observed and exclusion limits on the parameters $|U_{e4}|^2$ and $m_4$ are obtained. 
Our best sensitivity is for masses around 20~eV, excluding $|U_{e4}|^2 \gtrsim 2\cdot10^{-2}$. Our result improves with respect to bounds set by the previous direct kinematic experiments Mainz and Troitsk. This search is complementary to reactor oscillation experiments and improves their constraints for $\Delta{m}^2_{41} \gtrsim $ 10~eV$^2$, excluding also a sizable fraction of the allowed parameters of the reactor and gallium anomalies. 
Our result shows the potential of KATRIN to probe the sterile neutrino hypothesis with the data collected for determining the active neutrino mass~\mnue. KATRIN will significantly improve its statistics in the next 5~years and further reduce its systematics and background. A large fraction of the reactor and gallium anomalies' region of interest will be probed to shed light on the existence of sterile neutrino admixtures.

\begin{acknowledgments}
We acknowledge the support of Helmholtz Association, Ministry for Education and Research BMBF (5A17PDA, 05A17PM3, 05A17PX3, 05A17VK2, and 05A17WO3), Helmholtz Alliance for Astroparticle Physics (HAP),  Helmholtz Young Investigator Group (VH-NG-1055), and Deutsche Forschungsgemeinschaft DFG (Research Training Groups GRK 1694 and GRK 2149, and Graduate School GSC 1085 - KSETA) in Germany; Ministry of Education, Youth and Sport (CANAM-LM2011019, LTT19005) in the Czech Republic; Ministry of Science and Higher Education of the Russian Federation under contract 075-15-2020-778; and the United States Department of Energy through grants  DE-FG02-97ER41020, DE-FG02-94ER40818, DE-SC0004036, DE-FG02-97ER41033, DE-FG02-97ER41041, DE-AC02-05CH11231, DE-SC0011091, and DE-SC0019304, and the National Energy Research Scientific Computing Center.
\end{acknowledgments} 

\bibliographystyle{bst/apsrev4-2}
\bibliography{ksn1}

\end{document}